\documentclass[aps,nofootinbib,superscriptaddress]{revtex4}
\pdfoutput=1
\usepackage{latexsym, graphicx}
\usepackage{amsmath,amssymb,amsthm}
\usepackage[pdftex,colorlinks]{hyperref}

\newcommand{\mtiny}[1]{\mbox{\tiny$#1$}}
\newcommand{\bra}[1]{\left\langle #1 \right|}
\newcommand{\ket}[1]{\left| #1 \right\rangle}

\theoremstyle{definition}

\theoremstyle{definition}

\numberwithin{equation}{section}

\begin{document}

\title{The Rotating-Wave Approximation: Consistency and Applicability from an Open Quantum System Analysis}

\author{Chris Fleming}
\email{chris_h_fleming@yahoo.com}

\author{N. I. Cummings}
\email{nickc@umd.edu}
\affiliation{Joint Quantum Institute and Department of Physics, University of Maryland, College Park, Maryland 20742-4111, USA}

\author{Charis Anastopoulos}
\email{anastop@physics.upatras.gr}
\affiliation{Department of Physics, University of Patras, 26500 Patras, Greece}

\author{B. L. Hu}
\email{blhu@umd.edu}
\affiliation{Joint Quantum Institute and Department of Physics, University of Maryland, College Park, Maryland 20742-4111, USA}
\affiliation{Maryland Center for Fundamental Physics, University of Maryland, College Park, Maryland 20742-4111, USA}

\date{March 8, 2010}

\begin{abstract}

We provide an in-depth and thorough treatment of the validity of the
rotating-wave approximation (RWA) in an open quantum system. We find
that when it is introduced after tracing out the environment, all
timescales of the open system are correctly reproduced, but the
details of the quantum state may not be.  The RWA made before the trace   
is more problematic: it results in incorrect values for
environmentally-induced shifts to system frequencies, and the
resulting theory has no Markovian limit.  We point out that great
care must be taken when coupling two open systems together under the      
RWA. Though the RWA can yield a master equation of Lindblad form
similar to what one might get in the Markovian limit with white
noise, the master equation for the two coupled systems is not a           
simple combination of the master equation for each system, as is
possible in the Markovian limit. Such a naive combination yields inaccurate dynamics.
To obtain the correct master equation for the composite system a
proper consideration of the non-Markovian dynamics is required.
\end{abstract}

\maketitle


\section{Introduction}
\subsection{Background}

The rotating-wave approximation (RWA) is used in many places in the
study of open quantum systems, particularly in the field of quantum
optics (see for example
\cite{ref:Breuer_2002,ref:Walls_Milburn_1995,ref:Carmichael_1999}),
but the validity of the approximation is treated in depth far less
often. There are actually two distinct rotating-wave approximations
both in widespread use : 1) the `pre-trace' (preT) RWA,  which
consists of modifying the interaction Hamiltonian by dropping the
so-called counter-rotating terms that are quickly oscillating in the
Dirac picture; and 2) the `post-trace' (posT) RWA, which is obtained
by neglecting terms in the master equation for the reduced density
matrix that are quickly oscillating in the Dirac picture (see, e.g.,
\cite{ref:Walls_Milburn_1995,ref:Carmichael_1999} and
\cite{ref:Breuer_2002,ref:Agarwal_1974} respectively).
Agarwal has carried out a systematic study
\cite{ref:Agarwal_1971,ref:Agarwal_1973,ref:Agarwal_1974}
differentiating between these two RWAs and addressed their validity
for atom-field interactions and spontaneous  emission processes.
More recently, various authors have claimed some features of the RWA
that may limit its applicability, which we will now discuss.

The most widely acknowledged problem with the preT RWA seems to be
that it yields incorrect frequency shifts in the atomic energy
levels, so that it is not suitable for calculating environmentally
induced level shifts or induced cooperative frequency shifts
\cite{ref:Agarwal_1971,ref:Agarwal_1973,ref:Agarwal_1974}.  West and
Lindenberg \cite{ref:west_1984} found  that the reduced system
dynamics obtained from the pre-trace RWA  does not have a Markovian
limit \footnote{These authors also claim that there is no
fluctuation-dissipation theorem for this model, a statement that the
present authors cannot agree with.}.  Finally, Ford and O'Connell
\cite{ref:ford_1997} have raised concerns that in general the total
Hamiltonian  obtained by the preT RWA does not have a spectrum which
is bounded below, and they suggest that this limits the
applicability of the approximation to first-order transition
amplitudes.

Other authors have raised a very different sort of concern about the
preT-RWA Hamiltonian for coupling of a localized system to a
quantized field, that it may produce spurious causality violation in
the calculations.  Consider, for example, a two-level atom in the
dipole approximation interacting with  the electromagnetic field.
The multipolar form of the Hamiltonian for this interaction is
\begin{equation}
\boldsymbol{H} = \boldsymbol{\vec{d}} \cdot
\boldsymbol{\vec{D}(\vec{R})} = \sum_{\vec{k}} \sum_{s=1}^2
\sqrt{\frac{2\pi \varepsilon_{\vec{k}}}{V}} \left( \vec{d} \cdot
\vec{e}_{\vec{k},s} \right) \left(\boldsymbol{\sigma}_+ e^{\imath
\omega t} + \boldsymbol{\sigma}_- e^{-\imath \omega t} \right)
\left( \boldsymbol{a}_{\vec{k},s} e^{-\imath \varepsilon_{\vec{k}}
t} - \boldsymbol{a}_{\vec{k},s}^\dagger e^{\imath
\varepsilon_{\vec{k}} t} \right)
\end{equation}
in the Dirac picture, and in the preT RWA the terms with frequency
$\omega+\varepsilon_{\vec{k}}$ would be neglected.  However, with
these terms dropped the interaction can no longer be expressed in
terms of the local field variable $\boldsymbol{\vec{D}(\vec{R})}$
\cite{ref:ClSi98}.  Indeed, a numerical study of a three-atom problem
\cite{ref:dolce_2006} found that noncausal terms appear when the
preT RWA is used, unless one makes the \emph{ad hoc} modification of
extending frequency integrals to $-\infty$. The
preT RWA may then misrepresent the effects of retarded propagation in
the electromagnetic field, which suggests problems with causality in
the study of multipartite systems.

Moreover, the Glauber detector model \cite{ref:Glauber}, long used
 in photodetection theory and quantum optics, uses the preT RWA, and
 it might give rise to quantum correlations between spacelike separated events that do
 not represent the effects of actual entanglement. The effective
 status of preT RWA in Glauber's theory is debated:
 some authors have shown that photodetection probabilities at short
  times appear to violate causality \cite{ref:Tatarski}, and modifications to Glauber's photodetection theory have been suggested \cite{ref:Tatarski, ref:Fleisch98}, while others  indicate that
    a different form of the RWA in photodetection theory can
    guarantee causality \cite{ref:Milonni95}.
  Our interest in this problem partly arose
from finding how the imposition of preT RWA affects the range of
validity of results from the calculation of the entanglement
dynamics of two atoms interacting with a common quantum field at
large atomic separation \cite{ref:ASH}.

Some form of the RWA is also invoked in discussions of open quantum
systems in the context of quantum optics and atom-field interactions.
Indeed, the standard derivation of the Born-Markov master equation in general requires an RWA to render it in \emph{Lindblad form} \cite{ref:Breuer_2002},
thus providing a completely-positive dynamical map (for all states at all times) as is conveniently assumed in many quantum information theory discussions.

With these two distinct RWAs in widespread use while some open
questions  remain about their limitations and  fallacies, we find it
useful to carry out a systematic analysis of the consistency and
applicability of the   RWAs in the modern language of open quantum
systems.  This would be fitting in view of the fact that researchers
today have to tackle problems beyond those of level population and
dissipation rates to deal with more subtle issues such as
quantum decoherence and entanglement dynamics and perform more
demanding tasks such as quantum state tomography and engineering.

\subsection{Main Results}
This major findings in this work are:
\begin{enumerate}
\item The RWA may be sufficient or insufficient depending on what information is desired about the system.
For the perturbative relaxation rates either the pre-trace or post-trace RWA is sufficient. To obtain the environmentally induced shifts in system frequencies, only the post-trace RWA is sufficient.  In order to get more detailed information about the evolution of the quantum state and the asymptotic steady state neither RWA is sufficient in general.
\item Even in a system where the underlying environmental noise is colored, the RWA can yield a Lindblad-form master equation similar to what one might get in the Markovian limit with white noise.  However, when two such systems are coupled together, one cannot obtain the correct dynamics by naively adding together the dissipative terms of the master equations, as would be possible in the Markovian limit.  While the master equation obtained by such a naive addition has a mathematically valid form, it does not generally yield the dynamics of the system obtained from a microscopic derivation of the master equation for the combined system (even after the RWA). There are sufficiently simple systems and bath correlations for which a less judicious application of the RWA can nevertheless produce an adequate master equation.
However, this is no guarantee for validity in more complex situations.
\item The pre-trace RWA does not in general have a Markovian limit.
\end{enumerate}

This paper is organized as follows:
In Sec.~\ref{sec: closedRWA} we review the RWA as it applies to interactions within a closed system.
In Sec.~\ref{sec: post_traceRWA} we discuss the consistency and applicability of the post-trace RWA.
In Sec.~\ref{sec: pre_traceRWA} we discuss the same for the pre-trace RWA.
In the last section we conclude with some discussions.

\section{The Rotating-Wave Approximation in Closed Systems}

\label{sec: closedRWA}
We first examine the rotating-wave approximation as ordinarily applied to a closed system consisting of several interacting subsystems. Its wave function $\ket{\psi}$ evolves as
\begin{eqnarray}
\frac{d}{dt} \ket{\psi} &=& -\imath\,  \mathbf{H} \ket{\psi} \, ,
\end{eqnarray}
under the total Hamiltonian $\mathbf{H} \equiv \mathbf{H}_0 + \mathbf{H}_1$ where $\mathbf{H}_0$ represents the sum of all uncoupled subsystems and $\mathbf{H}_1$ represents the part from the subsystems coupling.
One seeks to solve the eigenvalue problem
\begin{eqnarray}
\mathbf{H} \, \ket{\omega} &=& \omega \, \ket{\omega} \, ,
\end{eqnarray}
for eigenstates $\ket{\omega}$ with eigenvalues $\omega$ by perturbing off the free eigensystem
\begin{eqnarray}
\omega_i &=& \omega_i^{(0)} + \delta\omega_i^{(1)} + \cdots \, , \\
\ket{\omega_i} &=& \ket{\omega_i^{(0)}} + \ket{\delta\omega_i^{(1)}} + \cdots \, ,
\end{eqnarray}
where $\omega_i^{(0)}$ are the eigenvalues of the uncoupled system.
The \emph{non-degenerate} first-order corrections are then
\begin{eqnarray}
\delta\omega_i^{(1)} &=& \bra{\omega_i^{(0)}} \mathbf{H}_1 \ket{\omega_i^{(0)}} \, , \label{eq:pfreqCS} \\
\left\langle \omega_j^{(0)} \Big{|} \delta\omega_i^{(1)} \right\rangle &=& \frac{ \bra{\omega_j^{(0)}} \mathbf{H}_1 \ket{\omega_i^{(0)}} }{\omega_i - \omega_j} \,\, (\omega_i \neq \omega_j) \, , \label{eq:pbasisCS}
\end{eqnarray}
and for the \emph{degenerate} corrections one must find the correct linear combination of degenerate states $\ket{f}$,
which exist solely in the degenerate subspace $\ket{\omega_{d_i}^{(0)}}$ wherein $\mathbf{H}_0 \ket{\omega_{d_i}^{(0)}} = \omega_{d} \ket{\omega_{d_i}^{(0)}}$.
$\ket{f}$ are the eigenstates of the degenerate interaction $\mathbf{H_{d}}$
\begin{equation}
\mathbf{H_{d}} \ket{f} = f \, \ket{f} \, ,
\end{equation}
which possesses matrix elements
\begin{equation}
\left(H_{d}\right)_{ij} = \bra{\omega_{d_i}^{(0)}} \mathbf{H}_1 \ket{\omega_{d_j}^{(0)}} \, , \label{eq:degcharCS}
\end{equation}
This kind of analysis should also extend to \emph{nearly-degenerate} subspaces where the basis corrections in Eq.~\eqref{eq:pbasisCS} would nearly diverge.

In the rotating-wave approximation one only considers components of the interaction $\mathbf{H}_1$ which oscillate least rapidly in the interaction picture.
If these terms are stationary, e.g. at resonance, then they will include the correct non-degenerate first-order frequencies, Eq.~\eqref{eq:pfreqCS},
as well as the correct characteristic equation, Eq.~\eqref{eq:degcharCS}, which determines the degenerate first-order frequencies and zeroth-order energy states.
What the RWA generally neglects are the first-order basis corrections, Eq.~\eqref{eq:pbasisCS}.
If the RWA terms are non-stationary, then they will include the most nearly-degenerate first-order frequencies and also neglect their first order basis corrections.
Therefore the RWA has limited  correspondence to perturbation theory as long as all terms that are close to resonance have been retained in the interaction Hamiltonian.  If, however, the RWA is made such that there are neglected terms that are near resonance, then the correspondence fails even in the weak coupling regime (because the missing first-order basis corrections become large, and even the first-order eigenvalue corrections are inaccurate). Moreover, for subsystems with a multiplicity of timescales, there may be several near-resonance frequencies for which a proper application of RWA would have to take into account.

In the rest of this paper we consider open quantum systems and divide our attention between two cases: the post- and pre- trace RWA.

\section{The Post-Trace Rotating-Wave Approximation}
\label{sec: post_traceRWA}

An open-system is produced by coarse graining over the environmental
degrees of freedom and its dynamics is described by a master equation
governing the reduced density matrix (RDM) $\boldsymbol{\rho}$. In
the case that the effect of the environment can be represented by a
quantum white noise \footnote{Some authors refer to an environment or
its noise as Markovian or non-Markovian, but we maintain that the
terminology `Markovian' or `non-Markovian' be saved for the
description of stochastic \textit{processes}, not the nature of
noise.} the master equation has time-local coefficients and assumes a
\emph{Lindblad form}:
\begin{eqnarray}
\dot{\boldsymbol{\rho}} &=&  -\imath \left[ \mathbf{H} + \mathbf{V}, \boldsymbol{\rho} \right] + \boldsymbol{\mathcal{D}} \{ \boldsymbol{\rho} \} \, , \\
\boldsymbol{\mathcal{D}} \{ \boldsymbol{\rho} \} &\equiv& \sum_{nm} \mathcal{D}_{nm} \left( \mathbf{e}_n \, \boldsymbol{\rho} \, \mathbf{e}_m^\dagger -\frac{1}{2} \left\{ \mathbf{e}_m^\dagger \mathbf{e}_n , \boldsymbol{\rho} \right\} \right) \, , \label{eq:lindblad_form}
\end{eqnarray}
where $\mathbf{H}$ is the Hamiltonian of the free system and $\mathbf{V}$ is a correction introduced by the environment,
not necessarily to be renormalized in its entirety as it may contain nontrivial features such as diffusion components.
$\boldsymbol{\mathcal{D}}$ is the ``dissipator'' super-operator:
$\mathcal{D}_{nm}$ is a positive-definite and Hermitian coefficient matrix and $\mathbf{e}_n$ denotes a particular basis of representation for the dissipator.

More generally, one can motivate the Lindblad equation
\cite{ref:Lindblad,ref:Lindblad2} from Choi's theorem on
completely-positive maps \cite{ref:Choi}. The Lindblad equation
classifies all completely-positive master equations without time
dependence in their coefficients. Following the generalization of
Choi's theorem on merely Hermitian-preserving maps, one can quickly
prove that any Hermitian and trace-preserving master equation of
analogous time-local form must have a \emph{pseudo-Lindblad form}
with merely Hermitian $\boldsymbol{\mathcal{D}}$.

\subsection{General Features}
The master equation for the RDM of an open quantum system can be written in the form
\begin{equation}
\dot{\boldsymbol{\rho}} = \boldsymbol{\mathcal{L}}_0 \{
\boldsymbol{\rho} \} + \boldsymbol{\delta\mathcal{L}} \{
\boldsymbol{\rho} \} \, ,
\end{equation}
where
\begin{equation}
\boldsymbol{\mathcal{L}}_0 \{
\boldsymbol{\rho} \} = -\imath \left[ \mathbf{H} ,
\boldsymbol{\rho} \right] \,
\end{equation}
is the term that would already be present in the unitary dynamics of the closed system and $\boldsymbol{\delta\mathcal{L}}$ represents the dissipative corrections introduced by coupling to the environment.
Such open-system master equations can be derived perturbatively, in the system-environment interaction, with a variety of different techniques \cite{ref:kampen,ref:breuer01,ref:strunz}.
The rotating-wave approximation is often introduced, not in a purely perturbatively derived master equation, but in one derived via the Born-Markov approximation.
To second order in the system-environment interaction, the Born-Markov approximation is consistent with weak coupling perturbation, even well outside of the Markovian limit.
Therefore, given that the RWA will only be applicable to second order, it is of no consequence if one starts from the Born-Markov approximation or a more rigorous perturbative analysis.
One only has to keep in mind that RWA has no reliance upon any kind of Markovian approximation.

The post-trace (posT) RWA effectively consists of only
considering the parts of the super-operator
$\boldsymbol{\delta\mathcal{L}}$ which commute with the free system
propagation super-operator
\begin{eqnarray}
\mathbf{G}_0(t) \{ \boldsymbol{\rho} \} &=& e^{-\imath \mathbf{H} t}
\,  \boldsymbol{\rho} \, e^{+\imath \mathbf{H} t} \, .
\end{eqnarray}
If we consider evaluating our master equation coefficients in the energy basis,
 which in pseudo-Lindblad form amounts to resolving
\begin{eqnarray}
\mathcal{D}_{ki;jl} &=& \sum_{nm} \mathcal{D}_{nm} \bra{\omega_k}
\mathbf{e}_n \ket{\omega_i} \bra{\omega_j} \mathbf{e}_m^\dagger
\ket{\omega_l} \,
\end{eqnarray}
with $\ket{\omega_j}$ representing the system energy eigenstate with
frequency $\omega_j$, then the RWA essentially amounts to projecting
out the diagonal of this Hermitian matrix, i.e. terms with
$\omega_{k}-\omega_{i}=\omega_{j}-\omega_{l}$ or equivalently
$\omega_{i}-\omega_{j}=\omega_{k}-\omega_{l}$. For the perturbative
master equation to second order in the system-environment coupling,
which is assumed to be weak, these diagonal entries will settle to
positive values and therefore this projection yields a master
equation of the Lindblad form.  We will refer to the master equation
obtained this way as the RWA-Lindblad equation.

Such a Lindblad projection is only reasonable because the
system-environment coupling is assumed to be weak and the projection
is performed in the energy basis. As we will presently show, the
eigen-operators of the Liouvillian have the form $\ket{\omega_j}
\bra{\omega_k}$ plus corrections at second order in the
coupling strength so that the discrepancy introduced by the dropped
terms is small for sufficiently weak coupling.
The RWA-Lindblad equation is not fully
equivalent to the weak coupling master equation, but it generates an
evolution   which is very close to that of the weak coupling master
equation in a perturbative sense.

\subsubsection{Correspondence with Perturbation Theory}
\label{sec: pre_trace_perturb} Assuming the system-environment
coupling to be weak, one can solve an open-system master equation in
a perturbative fashion.
One seeks a solution to the eigenvalue problem
\begin{eqnarray}
\boldsymbol{\mathcal{L}} \, \boldsymbol{\sigma} &=& f \, \boldsymbol{\sigma} \, ,
\end{eqnarray}
for the matrices $\boldsymbol{\sigma}$ which would evolve with characteristic rate $f$,
given the perturbative expansions
\begin{eqnarray}
\boldsymbol{\mathcal{L}} &=& \boldsymbol{\mathcal{L}}_0 + \boldsymbol{\delta\mathcal{L}} \, , \\
\boldsymbol{\sigma}_{ij} &=& \ket{\omega_i}\!\!\bra{\omega_j} + \boldsymbol{\delta\sigma}_{ij} + \cdots \, , \\
f_{ij} &=& -\imath \omega_{ij} + \delta f_{ij} + \cdots \, ,
\end{eqnarray}
where the zeroth order terms are correctly set to match the free evolution of the system and $\omega_{ij}\equiv\omega_{i}-\omega_{j}$.
For Gaussian noise, which has very well behaved asymptotic properties, the perturbative corrections are at minimum second order in the system-environment interaction.
Therefore we will speak of all lowest order perturbative corrections as being second order.
The second-order constraint upon our eigenvalue problem is then
\begin{eqnarray}
\bra{\omega_k} \boldsymbol{\delta\mathcal{L}}\!\left\{ \ket{\omega_i}\!\!\bra{\omega_j} \right\} \ket{\omega_l}
&=& -\imath(\omega_{ij} - \omega_{kl}) \bra{\omega_k}
\boldsymbol{\delta\sigma}_{ij} \ket{\omega_l} + \delta f_{ij} \, \delta_{ij;kl} \, .
\end{eqnarray}
Evaluating the components of this equation yields the non-degenerate
corrections.
\begin{eqnarray}
\bra{\omega_k} \boldsymbol{\delta\sigma}_{ij} \ket{\omega_l} &=&
\frac{ \bra{\omega_k} \boldsymbol{\delta\mathcal{L}}\! \left\{
\ket{\omega_i}\!\!\bra{\omega_j} \right\} \ket{\omega_l}
}{-\imath(\omega_{ij}-\omega_{kl})} \mbox{~where~} \omega_{ij}
\neq \omega_{kl} \, , \label{eq:per_s} \\
\delta f_{ij} &=& \bra{\omega_i} \boldsymbol{\delta\mathcal{L}}\!
\left\{ \ket{\omega_i}\!\!\bra{\omega_j} \right\} \ket{\omega_j} \,
. \label{eq:per_f}
\end{eqnarray}
For $\boldsymbol{\sigma}_{ii}$  and $ f_{ii}$ the system is
degenerate, and one must solve the characteristic equation (Pauli master equation)
\begin{eqnarray}
\mathbf{W} \, \vec{\boldsymbol{\sigma}} &=& \delta f \, \vec{\boldsymbol{\sigma}} \, , \\
\bra{\omega_i} \mathbf{W} \ket{\omega_j} &=& \bra{\omega_i}
\boldsymbol{\delta\mathcal{L}}\! \left\{ \ket{\omega_j}\!\!\bra{\omega_j} \right\} \ket{\omega_i} \, , \\
\vec{\boldsymbol{\sigma}}_i & \equiv & \bra{\omega_i} \boldsymbol{\sigma} \ket{\omega_i} \, ,
\end{eqnarray}
for the branching under the perturbation.

The essential point  is that the perturbative corrections to the
eigenvalues are entirely captured by the post-trace RWA, while the
perturbative basis corrections are entirely neglected. The lack of
basis perturbation can be readily seen in the thermal state of the
system. The system evolving under the RWA-Lindblad equation will
relax into the thermal state described by the Boltzmann density
matrix  $\boldsymbol{\rho} \propto e^{-\beta \, \mathbf{H}}$. But
this is generally not how most systems would actually  thermalize in
a Hamiltonian formulation:  Given the full system + bath and
interaction Hamiltonian,  $\mathbf{H} + \mathbf{H}_\mathrm{B} +
\mathbf{H}_\mathrm{I}$, the system is expected to relax into a state
described by the reduced density matrix $\boldsymbol{\rho} \propto
\left\langle e^{-\beta (\mathbf{H} + \mathbf{H}_\mathrm{B} +
\mathbf{H}_\mathrm{I})} \right\rangle_{\!\mathrm{B}}$, assuming the
bath to be very large\footnote{That the entire system + environment appears thermalized in its reduction is proven to second order in Ref.~\cite{ref:QOS}.}.
This state typically only reduces to Boltzmann form in the limit of
zero system-bath interaction strength; however,
an infinitesimal interaction strength would imply an infinite relaxation time, so we cannot speak self-consistently of the system relaxing to the thermal state the RWA-Lindblad equation predicts.

This property suggests a limitation to the applicability of preT RWA in the study of entanglement dynamics.
In a multipartite system with components interacting only through the bath, this precludes the presence of asymptotic residual entanglement.
This is in contrast to the asymptotic behavior of bipartite systems in quantum Brownian motion--see, for example \cite{ref:PazRonc,ref:HuLin}.

\subsubsection{RWA Fails When Perturbation Theory Fails}
The second-order master equation should be valid when the second-order corrections provided by $\boldsymbol{\delta\mathcal{L}}$ are small as compared to the unperturbed dynamics generated by $\boldsymbol{\mathcal{L}}_0$.
Let us denote the strength of the dissipative corrections generically by the frequency $\gamma_\mathrm{D}$,
then the weak coupling condition is
\begin{eqnarray}
\gamma_\mathrm{D} & \ll & \min_{\omega_{ij} \neq 0} \left| \omega_{ij} \right| \, ,
\end{eqnarray}
as $\omega_{ij}$ corresponds to the eigenvalues of $\boldsymbol{\mathcal{L}}_0$ (not $\omega_{i}$ which corresponds to the eigenvalues of the unperturbed Schr\"{o}dinger equation).

The RWA-Lindblad equation does not directly correspond to the second-order master equation, but more correctly to the second-order solutions of said master equation.
For the second-order solutions to be valid the coupling must not only be weak in the above sense but also in the following sense
\begin{eqnarray}
\gamma_\mathrm{D} & \ll & \min_{\omega_{ij} \neq \omega_{kl}} \left| \omega_{ij} - \omega_{kl} \right| \, . \label{eq:weakc2}
\end{eqnarray}
which justifies the perturbative solutions.
One cannot have near degeneracy in the energy level splittings or the naive perturbative solutions, which the RWA-Lindblad equation corresponds to, will fail.
Perfect degeneracy is acceptable; the RWA-Lindblad equation retains these terms in the Pauli master equation for instance.
But near-degeneracy needs to be treated in a manner analogous to degeneracy; the nearly degenerate subspace should be diagonalized.
With the second-order master equation this is still possible, but
 with the RWA-Lindblad equation these terms have been discarded and
 one is left with an invalid master equation.
Thus, there can be situations where the weak-coupling condition is satisfied while the posT RWA condition is not.

This problem arises, for example, in cavity QED. A two-level atom of frequency $\Omega$ coupled to a resonant intracavity field mode will result in an energy spectrum of the composite system that has the form of the harmonic oscillator with each level split in two by the Rabi frequency $\sqrt{n}\Omega_{VR}$ of the dressed states,
\[
E_{n,\pm}=\hbar\left(\Omega n\pm\sqrt{n}\Omega_{VR}\right).
\]
If the intracavity field is coupled to the field outside the cavity, this becomes an open quantum system. If the intracavity field is coupled weakly enough to the atom, then the system will be in the weak-coupling regime of cavity QED and the vacuum Rabi frequency $\Omega_{VR}$ will be small compared to $\gamma_\mathrm{D}$.
In this case the posT RWA procedure does not, strictly speaking, apply as was noted by Scala et. al. \cite{ref:Turku_JCM}.

In such a case one can still do a partial RWA, neglecting terms that oscillate much faster than $\gamma_\mathrm{D}$ and keeping those that are slower.
This still leaves the master equation in pseudo-Lindblad form; however, assuming these timescales are sufficiently slow and the spectrum of environmental noise is sufficiently flat, one may be able to make an effective Markovian approximation for the remaining pseudo-Lindblad terms (even if one might not have been valid for original master equation due to the faster system dynamics that have been ignored in the posT RWA) to recover a Lindblad-form master equation. Scala et. al. argued this is the case for cavity QED with a low-temperature bath \cite{ref:Turku_BRWA}.

\subsubsection{Non-Markovian processes, nonlocal dissipation, colored noise}
\label{sec:system_spec} Before discussing the non-Markovian
properties of the RWA-Lindblad equation, it is beneficial to get a
clearer definition of \emph{non-Markovian}, as the term is applied in
a variety of incompatible ways in the  mathematics, physics and
chemistry literature. In an open quantum system perspective, the
environment acts as a stochastic source or noise, of both quantum and
classical (thermal) origins, influencing the system. This noise,
which imparts fluctuations and dissipation, can be multiplicative
when there is nonlinear interaction amongst the constituents in the
environment and colored when there is temporal correlation (memory)
which reflects the dynamical timescales of the environment (see
discussions e.g., in \cite{ref:QBM1,ref:CumHu}). When the effects of
a thermal reservoir with quantum mechanical degrees of freedom may be
represented by a quantum white noise source, the open quantum
system's noise kernel and damping kernel (the dissipation kernel's
anti-derivative) will have delta function correlations. Noise which
induces local dissipation will only necessarily have delta
correlations for the damping kernel. \emph{Non-Markovian} refers to
physical processes with memories. In an open-system framework the
influence of the coarse-grained environment can engender non-local
noise correlations even with local dissipation {\em or} in the high
temperature regime. White noise only enters at high temperature {\em
and} with local dissipation.

In the \emph{Markovian limit} the
timescales of the environment are taken to be much shorter than the timescales of the system.
Thus in an open-system perspective one cannot simplistically refer
to an environment as Markovian or non-Markovian in isolation from
the dynamics of the system which it influences but is measured in
reference to. A noise may appear `Markovian' only because its
characteristic time scales cannot be resolved by the system. Some
authors use the terminology `Markovian noise' or `non-Markovian
noise' to describe the \textbf{nature} of the noise.
 We prefer to refer the nature or properties of noise simply as white or  colored while reserve the terminology
 Markovian or non-Markovian for the description of
\textbf{processes}, without or with memory, as it should be.

Another common habit is the use of Markovian in reference to the master equation itself, if it is time-local and especially if it is additionally time-homogeneous.
This should more correctly be referred to as a \emph{Markovian representation} or simply avoided.
Markovian processes produce Markovian representations, but not all Markovian representations arise from Markovian processes.
A master equation which arises from a non-Markovian process, even if in Markovian representation, is not sufficient to generate the dynamics of multi-time correlations.
By definition, Markovian processes must give rise to a Quantum Regression Theorem in which higher order correlations can be determined from lower order correlations and ultimately everything can be resolved via the master equation.
But with a non-Markovian process there are corrections to the Quantum Regression Theorem\cite{ref:QRT1,ref:QRT2}.

\subsubsection{Constant-in-time, Time-local and Time-nonlocal Master Equations}

A master equation for the reduced density matrix is \emph{time-local} or \emph{convolutionless}
if the time derivative of the (reduced) density
matrix at a time $t$ is expressed only in terms of the (reduced)
density matrix at that time $\boldsymbol{\rho}(t)$ (as opposed to an
integro-differential equation depending on the past history of
$\boldsymbol{\rho}$). A master equation can be time-local and yet
non-Markovian. A well-known example is the HPZ master equation
\cite{ref:QBM1} where the coefficients are obtained from solutions
of ordinary differential equations, the non-Markovian features
manifest in the nonlocal dissipation arising from the back-action of
the environment with colored noise on the system.

The corresponding quantum Langevin equations (Heisenberg equations of
motion for system operators with the  reservoir operators integrated
out) will remain non-local in time. The class of time-local master
equations  is distinguished from the class of master equations with
constant coefficients such as the familiar Lindblad  master equation
on the one hand, and the time-nonlocal (integro-differential)
equations on the other hand.  The latter form is often encountered
in a projection operator formalism with little or no coarse-graining
  of the environment. (See, e.g., \cite{ref:Zwanzig}).

A non-Markovian master equation will exhibit memory in a manner
that is not naively apparent in its time-local representation. This
is true even of the RWA-Lindblad equation, despite the fact that it
is of Lindblad form as one would have in a Markovian limit.
Let us consider the master equations for the reduced density
matrices of the (open) systems A and B with Hamiltonians
$\mathbf{H}_\mathrm{A}, \mathbf{H}_\mathrm{B}$  each coupled to a
dissipative environment of equivalent influence. We then have the open system
\begin{eqnarray}
\dot{\boldsymbol{\rho}}_\mathrm{A} &=& -\imath[ \mathbf{H}_\mathrm{A} , \boldsymbol{\rho}_\mathrm{A} ] + \boldsymbol{\delta \mathcal{L}}_\mathrm{A} \{ \boldsymbol{\rho}_\mathrm{A} \} \, , \\
\dot{\boldsymbol{\rho}}_\mathrm{B} &=& -\imath[ \mathbf{H}_\mathrm{B} , \boldsymbol{\rho}_\mathrm{B} ] + \boldsymbol{\delta \mathcal{L}}_\mathrm{B} \{ \boldsymbol{\rho}_\mathrm{B} \} \, ,
\end{eqnarray}
where the Hamiltonians are those of the free systems and the
corrections to the Liouville operator are introduced via interaction
with the dissipative environment. In the Hamiltonian formalism one
can simply add two Hamiltonians and arrive at another Hamiltonian,
though one might be motivated to fix the energy spectrum through
renormalization. One cannot do this with non-Markovian Liouville
operators. E.g. given some coupling  $\mathbf{H}_\mathrm{AB}$
between subsystems one cannot simply add dissipative terms.
\begin{eqnarray}
\dot{\boldsymbol{\rho}}_\mathrm{A+B} & \neq & -\imath[ \mathbf{H}_\mathrm{A} +
 \mathbf{H}_\mathrm{B} + \mathbf{H}_\mathrm{AB} , \, \boldsymbol{\rho}_\mathrm{A+B} ] +
 \boldsymbol{\delta \mathcal{L}}_\mathrm{A} \{ \boldsymbol{\rho}_\mathrm{A+B} \} +
\boldsymbol{\delta \mathcal{L}}_\mathrm{B} \{ \boldsymbol{\rho}_\mathrm{A+B} \} \, , \label{eq:combo_master_eq}
\end{eqnarray}
The above (incorrect) master equation is in general completely
different from the correct open-system master equation
 derived from first principles. For non-Markovian processes,
the environmental contributions have a nontrivial dependence
(due to memory effects) upon the system's dynamics through their couplings. If
one changes the system Hamiltonian then one must also change the
environmental contributions to be compatible with the history these
new terms will create. This is how memory exhibits itself in a
time-local representation. Moreover, one must also take into account
whether or not the dissipative environments are separate or shared.
If the dissipative environment is shared then the two subsystems can
interact via environmental back-reaction. This effect is also missed
when simply combining the Liouville operators.

In a general non-Markovian master equation, the problems of the
above incorrect master equation would be readily apparent as it
would likely violate positivity, uncertainty, etc. Positivity
violation will not occur when adding RWA-Lindblad terms, but the
mistake has only become more subtle and therefore more dangerous.
The master equation might be completely positive, but it does not
correspond to the dynamics of the physical system considered. This
is the key point.

This issue has already been commented on in the context of cavity
QED. The often-used master equation includes the Hamiltonian for the
atom, intracavity field, and atom-field interaction, but the
dissipator used is exactly that of an empty cavity with dissipation
plus that of an atom spontaneously emitting into empty space,
so that the situation is just that depicted in
 Eq.~\eqref{eq:combo_master_eq}.  And, indeed, if one begins instead with the
atom-cavity system and derives the microscopic master equation using
the standard technique \cite{ref:Breuer_2002}, one finds that the
master equation has a different dissipative term
\cite{ref:Turku_JCM,ref:Turku_BRWA}. As explained in
\cite{ref:Turku_BRWA} if the spectrum of environmental noise is
sufficiently flat then the difference is suppressed, which explains
the success of the standard cavity QED master equation. But not so
otherwise, which is something often overlooked.

\subsection{Application to the Two-Level Atom}
Here we consider a two-level system with $\boldsymbol{\sigma}_z$
Hamiltonian and energy level splitting
 $\Omega$, bilinearly coupled to a thermal reservoir via a $\boldsymbol{\sigma}_x$ coupling.  This would,
 for example, model a two-level atom coupled to the electromagnetic field in the dipole approximation.
Denoting our reduced density matrix
\begin{eqnarray}
\boldsymbol{\rho} &=& \left[ \begin{array}{cc} \rho_{\mtiny{++}} & \rho{\mtiny{+-}} \\
\rho{\mtiny{-+}} & \rho{\mtiny{--}} \end{array} \right] \, ,
\end{eqnarray}
one can compute the second-order master equation \cite{ref:QOS} and
place it into the form
\begin{eqnarray}
\frac{d}{dt} \left[ \begin{array}{c} \rho_{\mtiny{++}}
\\\rho{\mtiny{--}} \end{array} \right] &=&
\frac{\Gamma}{\cosh\!\left( \frac{\Omega}{2T} \right)} \left[
\begin{array}{cc} -e^{+\frac{\Omega}{2T}} &
 +e^{-\frac{\Omega}{2T}} \\ +e^{+\frac{\Omega}{2T}} & -e^{-\frac{\Omega}{2T}} \end{array} \right] \left[
  \begin{array}{c} \rho_{\mtiny{++}} \\ \rho{\mtiny{--}} \end{array} \right] \, , \\
  \frac{d}{dt} \left[ \begin{array}{c} \rho_{\mtiny{+-}} \\ \rho{\mtiny{-+}} \end{array} \right]
&=& \left[ \begin{array}{cc} -\Gamma - \imath (\Omega - \delta \Omega) & +\Gamma + \imath \, \delta \Omega \\
+\Gamma - \imath \, \delta \Omega & -\Gamma +\imath(\Omega - \delta \Omega) \end{array} \right]
\left[ \begin{array}{c} \rho_{\mtiny{+-}} \\ \rho{\mtiny{-+}} \end{array} \right] \, ,
\end{eqnarray}
with decoherence rate (here also the half thermalization rate)
and energy level shift
\begin{eqnarray}
\Gamma &\equiv& \Gamma(\Omega) \, , \\
\delta \Omega &\equiv& \frac{2}{\pi} \int_0^\infty \!\! d\varepsilon \, \mathcal{P}\!
\left[ \frac{\Omega}{\varepsilon^2-\Omega^2} \right] \Gamma(\varepsilon) \, ,
\end{eqnarray}
in terms of the phenomenological decoherence rate function $\Gamma(\omega)$.
$\mathcal{P}$ denotes the Cauchy principal value which regulates contained poles from contributing to the integral.

This master equation and those that follow are exact to second order, only the coefficients have been allowed to relax to their asymptotic values.
The relaxation occurs quickly, within the system and bath timescales, as compared to their effect, which occurs in the coupling timescale.
Therefore, when considering properly correlated initial states which do not jolt, it is safe to consider this ``late-time'' regime.

In terms of the microscopically derived damping kernel $\tilde{\gamma}(\omega)$, the anti-derivative of the dissipation kernel, the
decoherence rate can be expressed
\begin{eqnarray}
\Gamma(\Omega) &=& \tilde{\gamma}(\Omega) \, \Omega \coth\!\left( \frac{\Omega}{2T} \right) \, .
\end{eqnarray}
Regardless of system-environment coupling, the damping kernel is
effectively constant for Ohmic coupling, which along with high
temperature is responsible for thermal white noise. For linear
coupling to the collective positions of a bath of harmonic
oscillators the dissipation kernel has no more temperature
dependence than the system-environment coupling itself.

The post-trace RWA here amounts to neglecting the dynamical
interaction between $\rho{\mtiny{+-}}$ and $\rho{\mtiny{-+}}$. To
second order in the coupling, the only effect of this is to neglect
a perturbative amount of phase  information pertaining to their
damped oscillations, i.e. the perturbative change of basis. The
asymptotic state works out to be exactly the same in either case.
Thus under the specific conditions leading to these results the
post-trace RWA can be viewed as largely acceptable and somewhat
innocuous.

\subsection{Application to Quantum Brownian Motion}
The exact Quantum Brownian Motion (QBM) master equation \cite{ref:QBM1} for the reduced dynamics of
an oscillator bilinearly coupled (position-position) to a thermal reservoir of harmonic oscillators can be written
\begin{eqnarray}
\dot{\boldsymbol{\rho}} &=& -\imath [ \mathbf{H}_\mathrm{R} , \boldsymbol{\rho} ] -
 \imath \Gamma [ \mathbf{x} , \{ \mathbf{p} , \boldsymbol{\rho} \} ] -
 M \, D_{pp} [ \mathbf{x} , [ \mathbf{x} , \boldsymbol{\rho} ]] -
 D_{xp} [ \mathbf{x} , [ \mathbf{p} , \boldsymbol{\rho} ]] \, ,
\end{eqnarray}
where $\mathbf{H}_\mathrm{R}$ denotes the system Hamiltonian with renormalized frequency coefficient,
$\Gamma$ the dissipation coefficient,
$D_{pp}$ and $D_{xp}$ the regular and anomalous diffusion coefficients.
This master equation will have a stationary limit if the noise correlation is not
excessively widespread in time, e.g. a regulated ohmic
coupling is perfectly suitable. To lowest order in the coupling, the
late-time expressions for these coefficients can be determined from
 the weak coupling master equation \cite{ref:QOS} to be
\begin{eqnarray}
\Gamma &=& \Gamma(\Omega) \, , \\
\Omega_\mathrm{R} &=& \Omega - \frac{2}{\pi} \int_0^\infty \!\! d\varepsilon \, \mathcal{P}\!
\left[ \frac{\varepsilon^2}{\varepsilon^2-\Omega^2} \right] \Gamma(\varepsilon) \, , \\
D_{pp} &=& \Gamma(\Omega) \, \Omega \, \coth\!\left( \frac{\Omega}{2T} \right) \, , \\
D_{xp} &=& + \frac{2}{\pi} \int_0^\infty \!\! d\varepsilon \, \mathcal{P}\!
\left[ \frac{1}{\varepsilon^2-\Omega^2} \right] \Gamma(\varepsilon) \, \varepsilon \coth\!
\left( \frac{\varepsilon}{2T} \right) \, ,
\end{eqnarray}
in terms of the phenomenological dissipation function, which is
proportional to the dissipation kernel at second order. In QBM with
linear coupling to the bath, the dissipation coefficient will always
have no more temperature dependence than the system-environment
coupling. Ohmic coupling here will imply a constant dissipation
function, though a cutoff is required for this theory. The frequency
shift will have a linear cutoff sensitivity which can be
renormalized, but the anomalous diffusion $D_{xp}$ has a logarithmic
cutoff sensitivity.

For this problem, the Fokker-Planck equation for the probability distribution function (Wigner function) W in a phase space representation presents a much cleaner picture with simple solutions
\cite{ref:QBM}.
\begin{eqnarray}
\dot{W} &=& \left\{ \boldsymbol{\nabla}^\mathrm{T} \, \boldsymbol{\mathcal{H}} \,
\mathbf{q} + \boldsymbol{\nabla}^\mathrm{T} \,
\mathbf{D} \, \boldsymbol{\nabla} \right\} W \, , \\
\mathbf{q} &=& [ x , p ]^\mathrm{T} \, , \\
\boldsymbol{\nabla} &=& \left[ \frac{\partial}{\partial x}
,\frac{\partial}{\partial p} \right]^\mathrm{T} \, ,
\end{eqnarray}
The matrices $\boldsymbol{\mathcal{H}}$ and $\mathbf{D}$ are the homogeneous and diffusion coefficient matrices respectively.
\begin{eqnarray}
\boldsymbol{\mathcal{H}} &=& \left[ \begin{array}{cc} 0 & -\frac{1}{M} \\ M \Omega_\mathrm{R}^2 & 2 \Gamma \end{array} \right] \, , \\
\mathbf{D} &=& \left[ \begin{array}{cc} 0 & -\frac{1}{2}D_{xp} \\ -\frac{1}{2}D_{xp} & M \, D_{pp} \end{array} \right] \, .
\end{eqnarray}
Do not confuse the homogeneous generator with the Hamiltonian; they
differ by some frequency renormalization and the dissipation
$\Gamma$. From hereon we will assume the system frequency to be
properly renormalized such that $\Omega_\mathrm{R}=\Omega$ in the
stationary limit. The diffusion matrix contains two components: the
regular diffusion $D_{pp}$ and an anomalous anti-diffusion $D_{xp}$
which keeps the position uncertainty insensitive to high frequency.

If one expresses the QBM master equation in terms of ladder
operators, the pseudo-Lindblad coefficient matrix can be calculated
to be
\begin{eqnarray}
\boldsymbol{\mathcal{D}} &=& \frac{1}{\Omega} \left[
\begin{array}{cc} D_{pp} - \Gamma \, \Omega & D_{pp} + \imath D_{xp}
\, \Omega \\ D_{pp} - \imath D_{xp} \, \Omega & D_{pp} + \Gamma \,
\Omega \end{array} \right] \, .
\end{eqnarray}
The rotating-wave approximation then constitutes projecting out the
diagonal of this matrix, which will be positive definite.
Transforming back into the phase space representation, the
Fokker-Plank coefficients become
\begin{eqnarray}
\boldsymbol{\mathcal{H}}_\mathrm{RWA} &=& \left[ \begin{array}{cc} \Gamma & -\frac{1}{M}  \\ M \Omega_\mathrm{R}^2 & \Gamma \end{array} \right] \, , \\
\mathbf{D}_\mathrm{RWA} &=& \left[ \begin{array}{cc} \frac{D_{pp}}{2M\Omega^2} & 0 \\ 0 & \frac{M D_{pp}}{2} \end{array} \right] \, .
\end{eqnarray}
The anomalous diffusion coefficient vanishes entirely while the
dissipation and regular diffusion coefficients are both broken in
half, with the missing half reappearing as an analogous coefficient
of the master equation.

The role of the homogeneous coefficients are to generate the
homogeneous propagator $e^{-t \boldsymbol{\mathcal{H}}}$. The RWA homogeneous
coefficients are just slightly off in both the oscillation rates and
phase; the dissipation rates are entirely correct. Compare the
characteristic frequencies of the two matrices
\begin{eqnarray}
h &=& \Gamma \pm \imath \sqrt{\Omega^2-\Gamma^2} \, , \\
h_\mathrm{RWA} &=& \Gamma \pm \imath \Omega \, .
\end{eqnarray}

The diffusion coefficients are relatively more mangled given that
the anomalous coefficient is entirely absent; despite what is amiss
in the early calculations of the weak coupling QBM master equation,
this coefficient does exist at lowest order in the
system-environment coupling. The effect of diffusion is only present
in the second cumulant or covariance of the Wigner function. For
this stationary master equation, the evolution of the covariance is
simply
\begin{eqnarray}
\boldsymbol{\sigma}(t) &=& e^{-t \boldsymbol{\mathcal{H}}} \left[
\boldsymbol{\sigma}(0) - \boldsymbol{\sigma}(\infty) \right] e^{-t
\boldsymbol{\mathcal{H}}^\mathrm{T}} + \boldsymbol{\sigma}(\infty) \, ,
\end{eqnarray}
where the stationary covariance is determined by the Lyapunov equation
\begin{eqnarray}
\boldsymbol{\mathcal{H}} \, \boldsymbol{\sigma}(\infty) +
\boldsymbol{\sigma}(\infty) \, \boldsymbol{\mathcal{H}}^\mathrm{T} &=& 2 \,
\mathbf{D} \, .
\end{eqnarray}
We can easily compare the stationary covariances.
\begin{eqnarray}
\boldsymbol{\sigma}(\infty) &=& \left[ \begin{array}{cc} \frac{1}{M\Omega^2} \left( \frac{1}{2\Gamma} D_{pp} - D_{xp} \right) & 0 \\ 0 & \frac{M}{2\Gamma} D_{pp} \end{array} \right] \, , \\
\boldsymbol{\sigma}_\mathrm{RWA}(\infty) &=& \left[ \begin{array}{cc} \frac{1}{M\Omega^2} \frac{1}{2\Gamma} D_{pp} & 0 \\ 0 & \frac{M}{2\Gamma} D_{pp} \end{array} \right] \, .
\end{eqnarray}
Amazingly the only difference in the stationary state will come from
the lack of an anomalous diffusion coefficient. This contribution
will ultimately be lower order in the coupling, due to the
$\Gamma^{-1}$ prefactor before $D_{pp}$, and therefore its absence
is acceptable perturbatively.

\section{The Pre-Trace Rotating-Wave Approximation}
\label{sec: pre_traceRWA}

\subsection{General Features}

\subsubsection{Inconsistency of Approximation}
Let us consider a bilinear interaction Hamiltonian $\mathbf{H}_\mathrm{I}$  between a system observable
$\mathbf{L}$ and the collective environment observable $\mathbf{l}$.
\begin{eqnarray}
\mathbf{H}_\mathrm{I} &=& \mathbf{L} \, \mathbf{l} \, ,
\end{eqnarray}
For each of these operators, assuming them to be completely
non-stationary, there is a \emph{gross} raising and lowering
decomposition $\mathbf{L}_\pm$ given by
\begin{eqnarray}
\mathbf{L}_+ &=& \sum_{i>j} \bra{\omega_i} \mathbf{L} \ket{\omega_j} \, \ket{\omega_i}\!\bra{\omega_j} \, , \\
\mathbf{L}_- &=& \sum_{i<j} \bra{\omega_i} \mathbf{L} \ket{\omega_j} \, \ket{\omega_i}\!\bra{\omega_j} \, ,
\end{eqnarray}
such that
\begin{eqnarray}
\mathbf{L} &=& \mathbf{L}_+ + \mathbf{L}_- \, , \\
\mathbf{J} &=& \imath(\mathbf{L}_+ - \mathbf{L}_-) \, , \\
\mathbf{L}_\pm^\dagger &=& \mathbf{L}_\mp \, ,
\end{eqnarray}
where $\mathbf{L}$ and $\mathbf{J}$ will be two relevant observables.
For position coupling with a harmonic oscillator the decomposition becomes
\begin{eqnarray}
\mathbf{L} &=& \mathbf{x} \, , \\
\mathbf{L}_+ &=& \frac{1}{\sqrt{2 M \Omega}} \mathbf{a}^\dagger \, , \\
\mathbf{L}_- &=& \frac{1}{\sqrt{2 M \Omega}} \mathbf{a} \, , \\
\mathbf{J} &=& \frac{1}{M \Omega} \mathbf{p} \, ,
\end{eqnarray}
and for $\boldsymbol{\sigma}_x$ coupling with a $\boldsymbol{\sigma}_z$ Hamiltonian (two-level system) we have
\begin{eqnarray}
\mathbf{L} &=& \boldsymbol{\sigma}_x \, , \\
\mathbf{L}_\pm &=& \frac{1}{2} \boldsymbol{\sigma}_\pm \, , \\
\mathbf{J} &=& -\boldsymbol{\sigma}_y \, .
\end{eqnarray}

Now consider coupling the system to an environment made of a large number of harmonic oscillators
in their collective positions.
Let us furthermore assume the system coupling is like that of the above harmonic oscillator or
two-level system such that it is characterized by a single frequency
$\omega$.
\begin{eqnarray}
\mathbf{l} &=& \sum_k c_k \, \mathbf{x}_k \, , \\
\mathbf{H}_\mathrm{I} &=& \sum_k \frac{c_k}{\sqrt{m_k
\varepsilon_k}}  \left\{ \left( \mathbf{L}_+ \, \mathbf{a}_k +
\mathbf{L}_- \, \mathbf{a}_k^\dagger \right) + \left( \mathbf{L}_+
\, \mathbf{a}_k^\dagger + \mathbf{L}_- \, \mathbf{a}_k \right)
\right\} \, ,
\end{eqnarray}
where $\mathbf{x}_k$ is the environment position operator with ladder operator $\mathbf{a}_k$, energy $\varepsilon_k$ and mass $m_k$.
In the interaction picture we have the interaction Hamiltonian
\begin{eqnarray}
\mathbf{H}_\mathrm{I}(t) &=& \sum_k \frac{c_k}{\sqrt{2 m_k \varepsilon_k}}
\left( \mathbf{L}_+ \, \mathbf{a}_k \, e^{+\imath(\omega-\varepsilon_k)t} +
\mathbf{L}_- \, \mathbf{a}_k^\dagger \, e^{-\imath(\omega-\varepsilon_k)t} \right) + \nonumber \\
&& \sum_k \frac{c_k}{\sqrt{2 m_k \varepsilon_k}}
\left( \mathbf{L}_+ \, \mathbf{a}_k^\dagger \, e^{+\imath(\omega+\varepsilon_k)t} + \mathbf{L}_- \, \mathbf{a}_k
 \, e^{-\imath(\omega+\varepsilon_k)t} \right) \, . \label{eq:preTTerms}
\end{eqnarray}
An often utilized pre-trace rotating-wave
approximation (preT RWA) is to neglect the second terms,
conventionally referred to as counter-rotating terms, as they are
deemed more rapidly oscillating than the first. However, this is only
true in a mode-by-mode comparison.
Keeping terms of frequency $|\omega-2\omega|=\omega$ while discarding terms of frequency
$|\omega+0|=\omega$ serves no good purpose. There is no a priori
sense in which this is an approximation at all, unless the only environment
modes which exist are near resonance.

A true bandwidth approximation which does what the preT RWA claims would instead modify the interaction Hamiltonian of Eq.~\eqref{eq:preTTerms} (before tracing out the environment) by neglecting all the "rapid" terms that oscillate with a frequency outside some frequency band $\Delta\omega$ in the interaction picture while retaining all the slower terms \footnote{Note that this "bandwidth" Hamiltonian does \emph{not} arise from restricting the field to some bandwidth of modes around the resonance frequency.}.  The resulting Hamiltonian would be

\begin{eqnarray}
\mathbf{H}_\mathrm{I}(t) &=& \sum_{\varepsilon_k = 0}^{\Delta\omega + \omega} \frac{c_k}{\sqrt{2 m_k \varepsilon_k}} \left( \mathbf{L}_+ \, \mathbf{a}_k \, e^{+\imath(\omega-\varepsilon_k)t} + \mathbf{L}_- \, \mathbf{a}_k^\dagger \, e^{-\imath(\omega-\varepsilon_k)t} \right) + \nonumber \\
&& \sum_{\varepsilon_k = 0}^{\Delta\omega - \omega} \frac{c_k}{\sqrt{2 m_k \varepsilon_k}}\left( \mathbf{L}_+ \, \mathbf{a}_k^\dagger \, e^{+\imath(\omega+\varepsilon_k)t} + \mathbf{L}_- \, \mathbf{a}_k \, e^{-\imath(\omega+\varepsilon_k)t} \right) \, \label{eq:bandwidth}
\end{eqnarray}
Note that if $\Delta\omega < \omega$, then the bandwidth approximated
Hamiltonian would have no counter-rotating terms .  Furthermore, if
the environment were such that all environmental frequencies
$\varepsilon_k$ lie in a band around resonance with $\left\vert
\varepsilon_k - \omega \right\vert < \omega$, then a bandwidth
approximation using this band would be equivalent to dropping all
counter-rotating terms.  However, in the general case the two
approximations are inequivalent, and simply dropping all
counter-rotating terms is inconsistent.

It is also important to note that if the bandwidth approximation of
Eq.~\eqref{eq:bandwidth} is performed with $\Delta\omega$ chosen such
that all near-degenerate terms are retained, then this is just the
sort of RWA we discussed in Sect.~\ref{sec: closedRWA}. The only
difference is that the environment is to be traced out at the end of
the calculation.  However, such a bandwidth approximation would
render the problem more difficult to solve than simply calculating a
full perturbative solution.

\subsubsection{Noise and the Markovian Limit}

The Hamiltonian obtained after RWA is not generally an approximation of the full interaction Hamiltonian for reservoirs.
It is nonetheless a linear Hamiltonian interaction with a thermal reservoir and will affect dissipation, decoherence, thermalization, etc.
Therefore it still possesses some of the same character to the original model.

Back in terms of observables, the RWA interaction Hamiltonian takes the form
\begin{eqnarray}
\mathbf{H}_\mathrm{I} &=& \frac{1}{2} \, \mathbf{L} \sum_k c_k \,
\mathbf{x}_k + \frac{1}{2} \, \mathbf{J} \sum_k c_k
\frac{\mathbf{p}_k}{m_k \omega_k} \, ,
\end{eqnarray}
and thus it describes a different but related set of system
variables coupled to a different, but related set of bath variables.
This results in two quantum noise sources
\begin{eqnarray}
\mathbf{l}_\mathrm{RWA} &=& \frac{1}{2} \sum_k c_k \, \mathbf{x}_k \, , \\
\mathbf{j}_\mathrm{RWA} &=& \frac{1}{2} \sum_k c_k \frac{\mathbf{p}_k}{m_k \omega_k} \, ,
\end{eqnarray}
which have not only autocorrelations in and of themselves but cross-correlations between
themselves.
Perhaps more clearly, if we consider the original damping kernel with
one $\mathbf{L}$-coupled source
\begin{eqnarray}
\tilde{\boldsymbol{\gamma}}(\omega) &=& \tilde{\gamma}(\omega)
\left[
\begin{array}{cc} 1 & 0 \\ 0 & 0 \end{array} \right] \, ,
\end{eqnarray}
then the RWA damping kernel with both $\mathbf{L}$ and $\mathbf{J}$-coupled sources becomes
\begin{eqnarray}
\tilde{\boldsymbol{\gamma}}_\mathrm{RWA}(\omega) &=&
\frac{\tilde{\gamma}(\omega)}{4}  \left[ \begin{array}{cc} 1 &
-\imath \, \mathrm{sign}(\omega) \\ + \imath \,
\mathrm{sign}(\omega) & 1 \end{array} \right] \, ,
\end{eqnarray}
with reference to the original damping kernel $\tilde{\gamma}(\omega)$.
The damping kernel, here in Fourier space, is defined (as in Ref.~\cite{ref:QOS})
as the anti-derivative of the multivariate dissipation kernel, which is itself the imaginary part of the multivariate noise correlation in the time domain.
The diagonal components come from the self-correlations $\langle \mathbf{l}(t) \, \mathbf{l}(\tau) \rangle_\mathrm{B}$ and $\langle \mathbf{j}(t) \, \mathbf{j}(\tau) \rangle_\mathrm{B}$,
while the off-diagonal components come from the cross-correlations $\langle \mathbf{l}(t) \, \mathbf{j}(\tau) \rangle_\mathrm{B}$ and $\langle \mathbf{j}(t) \, \mathbf{l}(\tau) \rangle_\mathrm{B}$.

There is a subtle pathology in the cross-correlations of these two
noise sources. The RWA interaction is an example of couplings to
\emph{different} kinds of bath observables with strong cross-coupling.
Such couplings do always not admit a Markovian limit.
The reason for this is because in addition to high temperature, the white noise limit also requires a local damping kernel, i.e. one constant in the Fourier domain.
This is not a problem with one noise source as one can typically choose an appropriate coupling, e.g. Ohmic, such that the damping kernel will work out to be local.
But with multivariate noise one must make all components of the damping tensor local, including new kinds of terms which arise from the cross-correlations.
Whether or not this is possible depends in part upon any relation between the self-correlations and the cross-correlations.

For the RWA damping tensor, if we make the diagonal components local
with what was Ohmic coupling, then the off-diagonal components will
appear highly non-local like sign$(\omega)$. But if we were to choose
a coupling as to make the off-diagonal components local, then the
diagonal components will necessarily be highly non-local. There is no
choice of coupling which can give us white noise. This problem with
the white noise limit of pre-trace RWA has been noted before
\cite{ref:west_1984}. 

\subsubsection{Correspondence with Perturbation Theory}
The perturbative correspondence between the pre-trace RWA and the
original model is a bit more complicated to demonstrate.
Let us start with the second-order corrections for our simple separable
coupling without any sort of RWA following Ref.~\cite{ref:QOS}.
\begin{eqnarray}
\bra{\omega_k} \boldsymbol{\delta\mathcal{L}}\!\left\{ \ket{\omega_i}\!\! \bra{\omega_j} \right\} \ket{\omega_l} &=& \bra{\omega_k} \mathbf{L} \ket{\omega_i} \left[ A(\omega_{ki})+\bar{A}(\omega_{lj}) \right] \bra{\omega_j } \mathbf{L} \ket{\omega_l} \notag \\
&& -\delta_{lj} \sum_{h} \bra{\omega_k} \mathbf{L} \ket{\omega_h} A(\omega_{hi}) \bra{\omega_h} \mathbf{L} \ket{\omega_i}  \notag \\
&& -\delta_{ki} \sum_{h} \bra{\omega_j} \mathbf{L} \ket{\omega_h} \bar{A}(\omega_{hj}) \bra{\omega_h} \mathbf{L} \ket{\omega_l} \, ,
\end{eqnarray}
with late-time master equation coefficients
\begin{eqnarray}
A(\omega) & \equiv & \frac{1}{2} \tilde{\alpha}(\omega) - \frac{\imath}{2 \pi} \int_{-\infty}^{+\infty} \!\!\! d\varepsilon \, \mathcal{P}\!\left[ \frac{1}{\omega-\varepsilon} \right] \tilde{\alpha}(\varepsilon) \, , \\
\alpha(t) & \equiv & \left\langle \mathbf{l}(t) \, \mathbf{l}(0) \right\rangle_{\!\mathrm{B}} \, ,
\end{eqnarray}
where $\alpha$ is the quantum noise correlation for our
stationary bath. These corrections capture all of the second-order
relaxation rates,  perturbative frequency shifts, and basis
corrections.

As we have discussed, the post-trace RWA essentially considers
taking  only the diagonal entries where $\omega_{ki}=\omega_{lj}$,
and under appropriate conditions this is sufficient to reproduce all of the perturbative
frequency shifts and relaxation rates but not the basis corrections.
\begin{eqnarray}
\bra{\omega_i+\omega} \boldsymbol{\delta\mathcal{L}}\!\left\{ \ket{\omega_i}\!\!\bra{\omega_j} \right\} \ket{\omega_j + \omega} &=& \bra{\omega_i+\omega} \mathbf{L} \ket{\omega_i} 2 \, \mathrm{Re}[A(\omega)] \, \overline{ \bra{\omega_j+\omega } \mathbf{L} \ket{\omega_j} } \notag \\
&& -\delta_{0\omega} \sum_{\omega'}  A(\omega') \left| \bra{\omega_i + \omega'} \mathbf{L} \ket{\omega_i} \right|^2  \notag \\
&& -\delta_{0\omega} \sum_{\omega'}  \bar{A}(\omega') \left| \bra{\omega_j + \omega'} \mathbf{L} \ket{\omega_j} \right|^2 \,.
\end{eqnarray}
One can see that the first terms, which directly correspond to the
pseudo-Lindblad dissipator, are now only determined by the real
part of $A(\omega)$ or the characteristic function of the noise
correlation $\tilde{\alpha}(\omega)$. This function is always
positive by Bochner's theorem.

The pre-trace RWA master equation has four related sets of terms
because of the two correlated noise sources. But as far as these
diagonal terms are concerned, which determine the perturbative
timescales, one can essentially consider a master equation of the
same form  but with the modified coefficients
\begin{eqnarray}
A_\mathrm{RWA}(\omega) &=& \frac{1}{2} \tilde{\alpha}(\omega)  -
\frac{\imath}{2 \pi} \int_{-\infty}^{+\infty} \!\!\! d\varepsilon
\frac{1+\mathrm{sign}(\omega) \, \mathrm{sign}(\varepsilon)}{2}
\mathcal{P}\!\left[ \frac{1}{\omega-\varepsilon} \right]
\tilde{\alpha}(\varepsilon) \, .
\end{eqnarray}
The real part, which determines the relaxation rates, remains
unchanged. But the imaginary part, which determines the energy level
shifts, is very different. So while the post-trace RWA can correctly produce all
of the perturbative timescales,  the pre-trace RWA can only produce
the relaxation rates, consistent with what has
been found in earlier specific cases
\cite{ref:Agarwal_1971,ref:Agarwal_1974}.

\subsubsection{Non-Markovian Nature of the Master Equation}
It is necessary to point out that, although the pre-trace RWA can often produce a master equation of Lindblad form, the coefficients are inherently non-Markovian. Even though the master equation is in a convolutionless form, the coefficients themselves contain integrals over the system's history alongside nonlocal correlations of the noise. As such, they cannot be universally applied to different systems (which would have different histories) even if one only wants the relaxation rates. This was easy to notice for the post-trace RWA as the correctly derived master equation coefficients would come out to be completely different. Here the reason is much the same.

For instance, let us consider an oscillator system with $\mathbf{x}$ coupling to the environment.
The accuracy of the pre-trace RWA decay rates stems from a correct $\mathbf{a}^\dagger$, $\mathbf{a}$ raising and lowering operator decomposition of $\mathbf{x}$.
This leads to a different but related model of environmental interaction with $\mathbf{x}$ and $\mathbf{p}$-coupled noise.
We have proven that the perturbative decay timescales will work out to be equivalent, but only by using the raising and lowering properties.
If we couple this oscillator to additional degrees of freedom in some larger system, then $\mathbf{a}^\dagger$ and $\mathbf{a}$ are no longer ensured to be raising and lowering operators for the new energy eigenstates of the system.  Once this criterion has been broken, the proof fails to apply and all coefficients of the misapplied master equation will likely be wrong.
A correct pre-trace RWA interaction would have to involve a raising and lowering operator decomposition which utilizes the full Hamiltonian of the larger system.

\subsection{Application to the Two-Level Atom}
Utilizing the second-order master equation, we find the RWA
interaction Hamiltonian yields
\begin{eqnarray}
\frac{d}{dt} \left[ \begin{array}{c} \rho_{\mtiny{++}} \\\rho{\mtiny{--}} \end{array} \right] &=& \frac{\Gamma}{\cosh\!\left( \frac{\Omega}{2T} \right)} \left[ \begin{array}{cc} -e^{+\frac{\Omega}{2T}} & +e^{-\frac{\Omega}{2T}} \\ +e^{+\frac{\Omega}{2T}} & -e^{-\frac{\Omega}{2T}} \end{array} \right] \left[ \begin{array}{c} \rho_{\mtiny{++}} \\ \rho{\mtiny{--}} \end{array} \right] \, , \\
\frac{d}{dt} \left[ \begin{array}{c} \rho_{\mtiny{+-}} \\ \rho{\mtiny{-+}} \end{array} \right] &=& \left[ \begin{array}{cc} -\Gamma - \imath (\Omega - \delta \Omega_\star) & +\Gamma + \imath \, \delta \Omega_\star \\ +\Gamma - \imath \, \delta \Omega_\star & -\Gamma +\imath(\Omega - \delta \Omega_\star) \end{array} \right] \left[ \begin{array}{c} \rho_{\mtiny{+-}} \\ \rho{\mtiny{-+}} \end{array} \right] \, ,
\end{eqnarray}
with new energy level shift
\begin{eqnarray}
\delta \Omega_\star &\equiv& \frac{1}{\pi} \int_0^\infty \!\! d\varepsilon \, \mathcal{P}\!\left[ \frac{1}{\varepsilon-\Omega} \right] \Gamma(\varepsilon) \, .
\end{eqnarray}
In addition to differing from the frequency shift without the RWA, it also contains a higher order cutoff sensitivity.
For approximately local dissipation, the sensitivity was logarithmic but is now linear.

\subsection{Application to Quantum Brownian Motion}
Again utilizing the second-order master equation, we  find the RWA
interaction Hamiltonian yields
\begin{eqnarray}
\dot{\boldsymbol{\rho}} &=& -\imath [ \mathbf{H}_\mathrm{R}^\star , \boldsymbol{\rho} ] - \imath \Gamma [ \mathbf{x} , \{ \mathbf{p} , \boldsymbol{\rho} \} ] - M \, D_{pp} [ \mathbf{x} , [ \mathbf{x} , \boldsymbol{\rho} ]] - D_{xp}^\star [ \mathbf{x} , [ \mathbf{p} , \boldsymbol{\rho} ]] \, ,
\end{eqnarray}
with new frequency shift and anomalous diffusion coefficient
\begin{eqnarray}
\Omega_\mathrm{R}^\star &=& \Omega - \frac{1}{\pi} \int_0^\infty \!\! d\varepsilon \, \mathcal{P}\!\left[ \frac{\varepsilon}{\varepsilon-\Omega} \right] \Gamma(\varepsilon) \, , \\
D_{xp}^\star &=& + \frac{1}{\pi} \int_0^\infty \!\! d\varepsilon \, \mathcal{P}\!\left[ \frac{\frac{1}{\Omega}}{\varepsilon-\Omega} \right] \Gamma(\varepsilon) \, \varepsilon \coth\!\left( \frac{\varepsilon}{2T} \right) \, .
\end{eqnarray}
For what was Ohmic coupling in the original model, the frequency
shift has  a different but still linear cutoff sensitivity. However
the anomalous diffusion coefficient now also has a linear cutoff
sensitivity. If the cutoff is very large, this could be very
problematic.


\subsection{A Multipartite Example}
Let us say that we have an array of, otherwise non-interacting, parallel q-bits all with $\boldsymbol{\sigma}_{\!z}$ Hamiltonians.
A simple dipole interaction can be represented with the bilinear interaction Hamiltonian
\begin{eqnarray}
\mathbf{H}_\mathrm{I} &=& \sum_n \boldsymbol{\sigma}_{\!x_n} \, \mathbf{l}_n
\end{eqnarray}
where $\boldsymbol{\sigma}_{\!x_n}$ is the $x$ spin component of the $n^\mathrm{th}$ qubit and $\mathbf{l}_n$ is its corresponding collective environment coupling.
The environmental coupling for a qubit at location $\mathbf{r}_n$ is
\begin{eqnarray}
\mathbf{l}_n &=& \sum_\mathbf{k} g_k \left\{ e^{+\imath \mathbf{k} \cdot \mathbf{r}_n} \, \mathbf{a}^\dagger_\mathbf{k} + e^{-\imath \mathbf{k} \cdot \mathbf{r}_n} \, \mathbf{a}_\mathbf{k} \right\} \, ,
\end{eqnarray}
with $g_k \propto 1/\sqrt{\varepsilon_k}$ for an electro-magnetic field environment such as discussed in Ref.~\cite{ref:Walls_Milburn_1995,ref:AnastopHu}.
The resultant damping kernel corresponding to the $\langle \mathbf{l}_n(t) \, \mathbf{l}_m(\tau) \rangle_\mathrm{B}$ correlation is
\begin{eqnarray}
\tilde{\gamma}_{nm}(\omega) &=& \tilde{\gamma}_{0} \, \mathrm{sinc}(r_{nm} \omega) \, ,
\end{eqnarray}
where $\mathbf{r}_{nm} = \mathbf{r}_n - \mathbf{r}_m$ and therefore the damping is Ohmic or local for the autocorrelations where $r_{nn}=0$.
The cross-correlations, which are strictly nonlocal, vanish in the limit of large distance separation.

The pre-trace RWA
interaction, which was considered  for two qubits in
Ref.~\cite{ref:ASH}, introduces the duplication of quantum noise
sources
\begin{eqnarray}
\mathbf{l}_n^\mathrm{RWA} &=& \frac{1}{2} \sum_\mathbf{k} g_k \left\{ e^{+\imath \mathbf{k} \cdot \mathbf{r}_n} \, \mathbf{a}^\dagger_\mathbf{k} + e^{-\imath \mathbf{k} \cdot \mathbf{r}_n} \, \mathbf{a}_\mathbf{k} \right\} \, , \\
\mathbf{j}_n^\mathrm{RWA} &=& \frac{\imath}{2} \sum_\mathbf{k} g_k \left\{ e^{+\imath \mathbf{k} \cdot \mathbf{r}_n} \, \mathbf{a}^\dagger_\mathbf{k} - e^{-\imath \mathbf{k} \cdot \mathbf{r}_n} \, \mathbf{a}_\mathbf{k} \right\} \, ,
\end{eqnarray}
and one must now consider the correlations between all such operators.
The resultant damping kernels can be organized
\begin{eqnarray}
\left[ \begin{array}{cc} \tilde{\gamma}_{l_n l_m}^\mathrm{RWA}(\omega) & \tilde{\gamma}_{l_n j_m}^\mathrm{RWA}(\omega) \\ \tilde{\gamma}_{j_n l_m}^\mathrm{RWA}(\omega) & \tilde{\gamma}_{j_n j_m}^\mathrm{RWA}(\omega) \end{array} \right]
&=&
\frac{\tilde{\gamma}_{nm}(\omega)}{4}  \left[ \begin{array}{cc} 1 & -\imath \, \mathrm{sign}(\omega) \\ + \imath \, \mathrm{sign}(\omega) & 1 \end{array} \right] \, ,
\end{eqnarray}
with reference to the original damping kernel $\tilde{\gamma}_{nm}(\omega)$.
The scenario is much the same.
The damping rates will be correct, while the frequency shifts and basis corrections (including asymptotic entanglement) will be incorrect.
There is no longer a white noise limit, even when the qubits are distantly separated.

However this remains a fairly reasonable physical theory, as the RWA interaction itself was fairly reasonable.
The cross-correlations between \emph{different} qubits still vanishes for large separations.
The second-order master equation, being determined by the second-order cumulants or two-time correlations, will therefore reduce to that of qubits coupled to independent environments in the large separation limit.

\section{Discussion}
\label{sec:discussion}

In this work we have systematically examined the rotating-wave
approximation by using a master equation for an open quantum system
weakly-coupled to a general environment.  There are, in fact, two
distinct rotating-wave approximations: The pre-trace RWA is an
approximation performed on the interaction Hamiltonian before the
environment is traced out which yields a somewhat modified
Hamiltonian dynamics from which the reduced dynamics can be derived.
The post-trace RWA is performed on the master equation for the
reduced density matrix after the environment has already been traced
out. In \cite{ref:QOS} we have derived the master equation for a
system with a discrete energy spectrum weakly coupled to a
stationary quantum bath. Using this general framework we have
compared the master equation describing the dynamics of this open
system under the pre and post-trace RWA to the dynamics without
making such an approximation. We have specifically addressed the
master equations for a two-level system and for a linear oscillator,
two models in which the RWA is often invoked.

We find the post-trace RWA to be more innocuous than the pre-trace RWA.
It can be seen as an often reasonable approximation in which the full weak-coupling Liouvillian (which is time-local and of pseudo-Lindblad form) is projected onto a Lindblad-form Liouvillian.
We call the resulting master equation the RWA-Lindblad equation.
We find that for a general open quantum system the post-trace RWA will yield exactly the same timescales as perturbative solutions of the weak-coupling master equation.
The perturbative corrections to eigen-operators of the Liouvillian are neglected in the RWA-Lindblad equation, so the predicted quantum state will differ.
In particular, the steady state solution of the RWA-Lindblad master equation will differ from the true steady state by an amount that is perturbative in the coupling.
These results are consistent with what Agarwal found \cite{ref:Agarwal_1971,ref:Agarwal_1973,ref:Agarwal_1974} for the two-level atom and the linear oscillator in the Born-Markov approximation.
One context in which the discrepancy in the steady  state could be important is examining the late-time behavior of  entanglement dynamics at low temperature.
When the system is bipartite and  the ground state is separable, the RWA-Lindblad equation will give an asymptotically separable state, whereas the weak-coupling master equation leaves open the possibility of asymptotic entanglement.
In view of our findings, we can say that generally the post-trace RWA is suitable if one only wants the perturbative timescales of the dynamics,
but it may not be appropriate if one wants more detailed information about the quantum state of the system, as it misses some corrections introduced by the coupling to the environment,
and it will also not be appropriate when perturbative timescales fail, i.e. for near-resonance in the energy level splittings.

We find the pre-trace RWA to be more problematic.   When the
environment contains many frequencies with a spread comparable to
the frequencies of the system, then the pre-trace RWA in general
does not provide a faithful representation of the true solution.  We
also find that the pre-trace RWA results in two strongly correlated
sources of environmental noise that together have no Markovian
limit.  The cross-correlations between the noise sources are such
that if the autocorrelations are white then the cross-correlations
are strongly colored.  This issue has been noticed before
\cite{ref:west_1984}.  Finally, we have shown that, unlike the
post-trace RWA, the pre-trace RWA in general does not correctly
obtain all perturbative timescales for the dynamics, yielding incorrect frequency
shifts. This finding based on a more extended theoretical framework
agrees with results obtained for specific cases studied before for
the two-level atom
\cite{ref:Agarwal_1971,ref:Agarwal_1973,ref:Agarwal_1974}.

We caution that the way the RWA is applied also matters.
For Markovian processes in closed and open systems, certain liberties can be taken with the master equation; terms can be mixed and matched rather haphazardly.
A Markovian process will generally produce a master equation of Markovian representation and Lindblad form.
The RWA can also produce a master equation of Markovian representation and Lindblad form, and therefore one might assume that these liberties can also be taken with the RWA-Lindblad master equation.
But this is not the case as the underlying stochastic process remains non-Markovian and the master equation
coefficients contain memory despite their Markovian representation.
A haphazard construction of RWA-Lindblad master equations for multipartite systems can produce
an evolution which is completely positive and yet totally unphysical.

The pre-trace RWA as applied to quantum Brownian motion is additionally
interesting because the interaction Hamiltonian in this case
coincides with the Gardiner-Collett Hamiltonian
\cite{ref:Gardiner_1985} used to model the coupling between the
intracavity field of a high-finesse electromagnetic cavity and
external field modes.  However, in this case the form of the
Hamiltonian is not the result of a rotating-wave approximation and
can be derived  \cite{ref:Dutra_2000} from a ``modes-of-the-universe''
approach for a cavity with a partially transmitting mirror in the
limit that transmission is weak.  So in this case the pre-trace RWA
Hamiltonian corresponds to the physical Hamiltonian of an actual
system, and the solutions of the master equation have relevance
directly, rather than as an approximation.

A single two-level atom is clearly a particularly simple quantum
system.  As such, some of the shortcomings of the RWA are obscured
in this case.  We find that the post-trace RWA gives the correct
equilibrium state for a thermal environment in this case, in
addition to the correct timescales, though it does miss some of the
corrections to the transient quantum  evolution that can be obtained
from the weak-coupling master equation without the RWA. Thus, if
one's theoretical investigations are limited to those features that
it captures correctly,  then the post-trace RWA  may be a suitable
approach.

Our general analysis provides a reasonably good
characterization of the validity of the RWA for constructing a
master equation; however, many  questions remain about its effects
on the internal dynamics of more complex systems, such as the
effect of either form of RWA on causality in multi-atom models with
spatial separation \cite{ref:dolce_2006} and its effect on entanglement dynamics  \cite{ref:ASH}.

\noindent{\bf Acknowledgment}  This work is supported partially by
NSF Grants PHY-0426696, PHY-0801368, DARPA grant
DARPAHR0011-09-1-0008 and the Laboratory of Physical Sciences.

\end{document}